\documentclass[10pt,letterpaper]{article}
\usepackage{opex3}
\usepackage{cite}

\usepackage{graphicx}
\usepackage{dcolumn}
\usepackage{bm}

\makeatletter
\def\@dotsep{4.5}
\makeatother

\begin{document}

\title{Cut-wire-pair structures as two-dimensional magnetic metamaterials}

\author{David A. Powell, Ilya V. Shadrivov, and Yuri S. Kivshar} 

\address{Nonlinear Physics Center, Research School of Physical
Sciences and Engineering\\
Australian National University, Canberra ACT 0200, Australia}

\email{david.a.powell@anu.edu.au}

\begin{abstract} We study numerically and experimentally magnetic metamaterials based on cut-wire pairs instead of split-ring resonators.  The cut-wire pair planar structure is extended in order to create a truly two-dimensional metamaterial suitable for scaling to optical frequencies.  We fabricate the cut-wire metamaterial operating at microwave frequencies with lattice spacing around 10\% of the free-space wavelength, and find good agreement with direct numerical simulations.  Unlike the structures based on split-ring resonators, the nearest-neighbor coupling in cut-wire pairs can result in a magnetic stop-band with propagation in the transverse direction.
\end{abstract}


Composite structures based on short pairs of wires have been proposed as a novel approach to achieving a magnetic response at optical frequencies~\cite{Podolskiy2002,Podolskiy2005a}.  The simple geometry of wire pairs means that such structures can be scaled to nanometer dimensions much more readily than those based on split-ring resonators.  The cut-wire structures also have a higher saturation frequency than split-ring resonators, meaning they can be scaled to shorter wavelengths before the dominance of electron kinetic energy terms in the inductance prevents further scaling.

So far all the experimental results for these cut-wire metamaterials have been reported for {\em planar structures} with a response engineered in a single direction only; in that sense all such structures can be termed as {\em one-dimensional}. This type of one-dimensional planar cut-wire-pairs structures has been experimentally tested for microwaves~\cite{souk_06,Guven2006,Lam_08} and has also been demonstrated to operate at optical frequencies~\cite{Shalaev2005,wegener_06}.

\begin{figure}[htb]
\begin{center}
 \includegraphics[width=0.8\columnwidth]{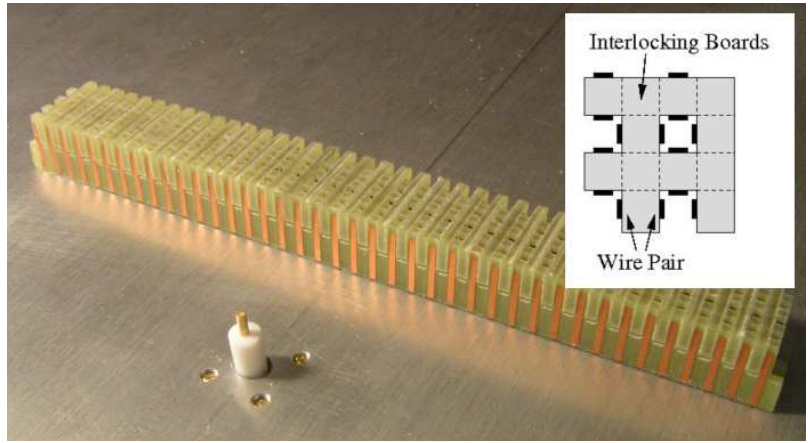}
\caption{Photograph of the two-dimensional cut-wire metamaterial placed
next to source antenna in the waveguide. Inset shows the arrangement of the wires.~\label{fig:structure}}
\end{center}
\end{figure}

However, the most exciting applications of engineered metamaterials such as cloaking and perfect lensing occur in two or three dimensional structures.  Thus, it is of a great importance to develop novel types of cut-wire structures with a magnetic response in more than one dimension and study their properties in detail.  In this paper, we suggest a novel geometry of {\em two-dimensional} cut-wire pair structures suitable for scaling to optical frequencies.  We fabricate a two-dimensional microwave metamaterial composed of pairs of cut wires designed to be isotropic within a plane and compare the responses of isotropic and anisotropic two-dimensional cut-wire metamaterials.

The magnetic metamaterial is constructed on 1.6mm FR4 with copper metallization.  The width of the wires is 1mm, the height 9mm, and the period 3mm.  The sample consists of 4x40 unit cells, each containing two resonant wire pairs aligned along the $X$ and $Y$ axes, aiming to yield an isotropic response in the $X-Y$ plane.  The boards are fabricated with vertical slots to enable them to be interlocked and are 10mm high.  A photograph of the fabricated structure in the measurement system is shown in Fig.~\ref{fig:structure}. The inset shows a plan of the wire arrangement.

\begin{figure}[htb]
\begin{center}
 \includegraphics[width=0.7\columnwidth]{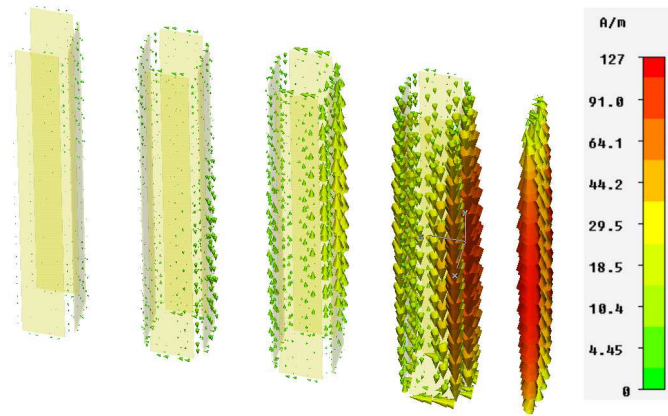}
\caption{Simulated current distribution within the stop-band
at 8GHz.  The wire pairs are formed by two wires with boards between them
(not shown), rather than those with a perpendicular wire pair between them.~\label{fig:currents}}
\end{center}
\end{figure}

The performance of this novel design is verified by numerical simulations using CST microwave studio.  Perfect electric conductor boundary conditions are applied to the top and bottom of the simulation domain to model the waveguide, and perfect magnetic conductor boundary conditions are applied to the sides to account for the periodicity.  The simulation domain, consisting of four layers, is illustrated in Fig.~\ref{fig:currents}, where the incident wave propagates from right to left, and the circuit boards are omitted for clarity.

\begin{figure}[htb]
\begin{center}
 \includegraphics[width=\columnwidth]{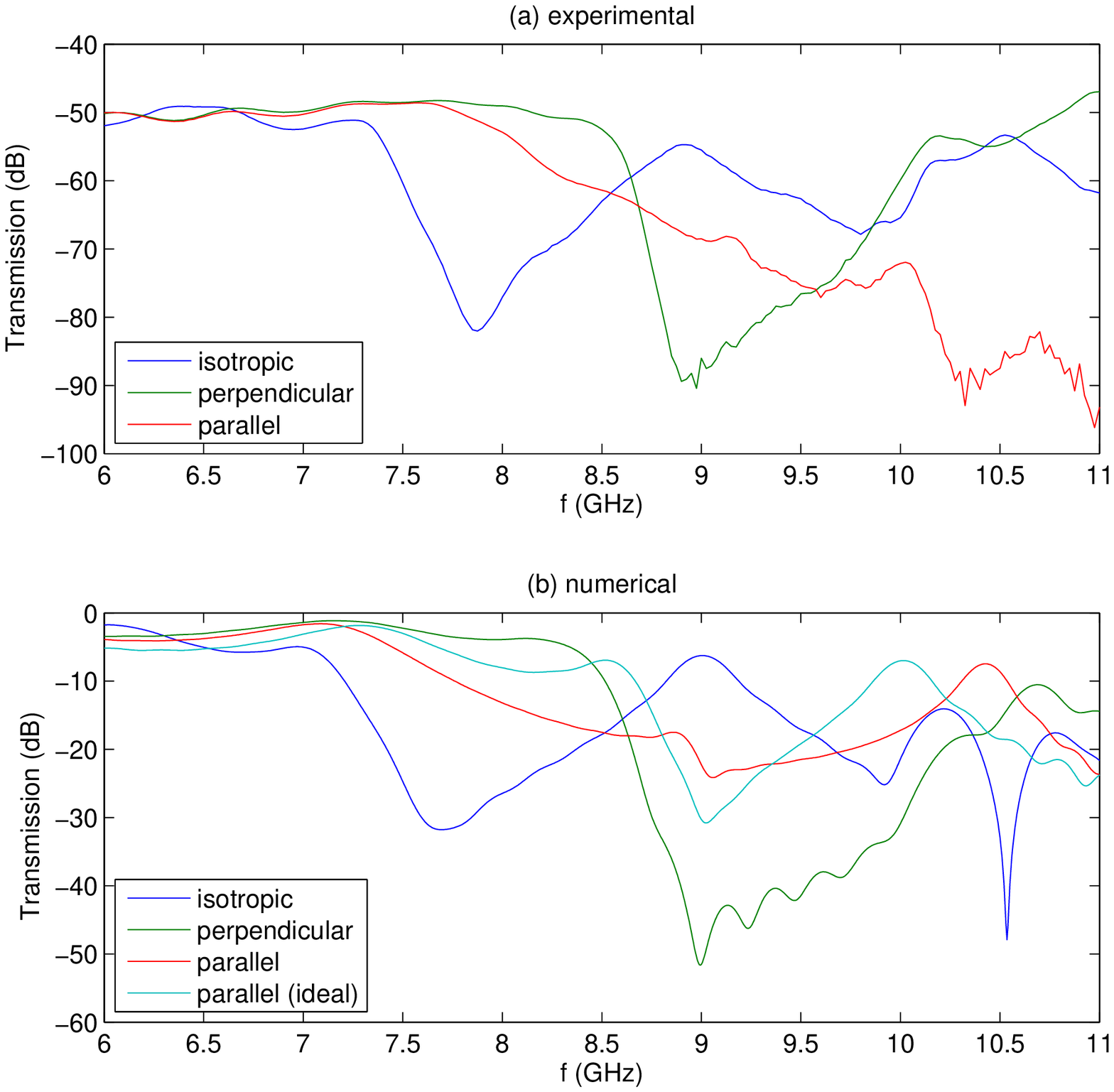}
\caption{(a) Measured and (b) simulated transmission spectra for
the isotropic case, as well as the two different anisotropic orientations of
boards.~\label{fig:transmission}}
\end{center}
\end{figure}

The experimental frequency response is obtained by placing the sample within a parallel plate waveguide with absorbing boundaries operating in its fundamental TEM mode, similar to that reported in Ref.~\cite{Schurig2006a}.  Transmission measurements are performed with a vector network analyzer using both a ``monopole antenna'' point source as shown in Fig.~\ref{fig:structure} and constructing a waveguide from absorbing foam to approximate plane wave excitation.  In both cases the results show the same key spectral features, however better agreement with simulation is achieved using the plane-wave type excitation.  Above 10GHz agreement is quite poor, most likely due to the finite width of the source beam, and the use of a point measurement at the output.

The experimental and numerical results are presented in the curves labeled ``isotropic'' in Fig.~\ref{fig:transmission}.  Good agreement is observed up to 10GHz, and a stop-band is clearly visible between 7.5-9GHz, which corresponds to a free space wavelength over 10 times the period of the composite structure.  In order to identify the type of this stop-band, the simulated surface currents are examined.  As can be seen in Fig.~\ref{fig:currents}, within each pair of wires the currents are out of phase with each other, indicating a current loop which couples to the incident magnetic field to create a magnetic stop-band.

\begin{figure}[htb]
\begin{center}
\includegraphics[width=0.75\columnwidth]{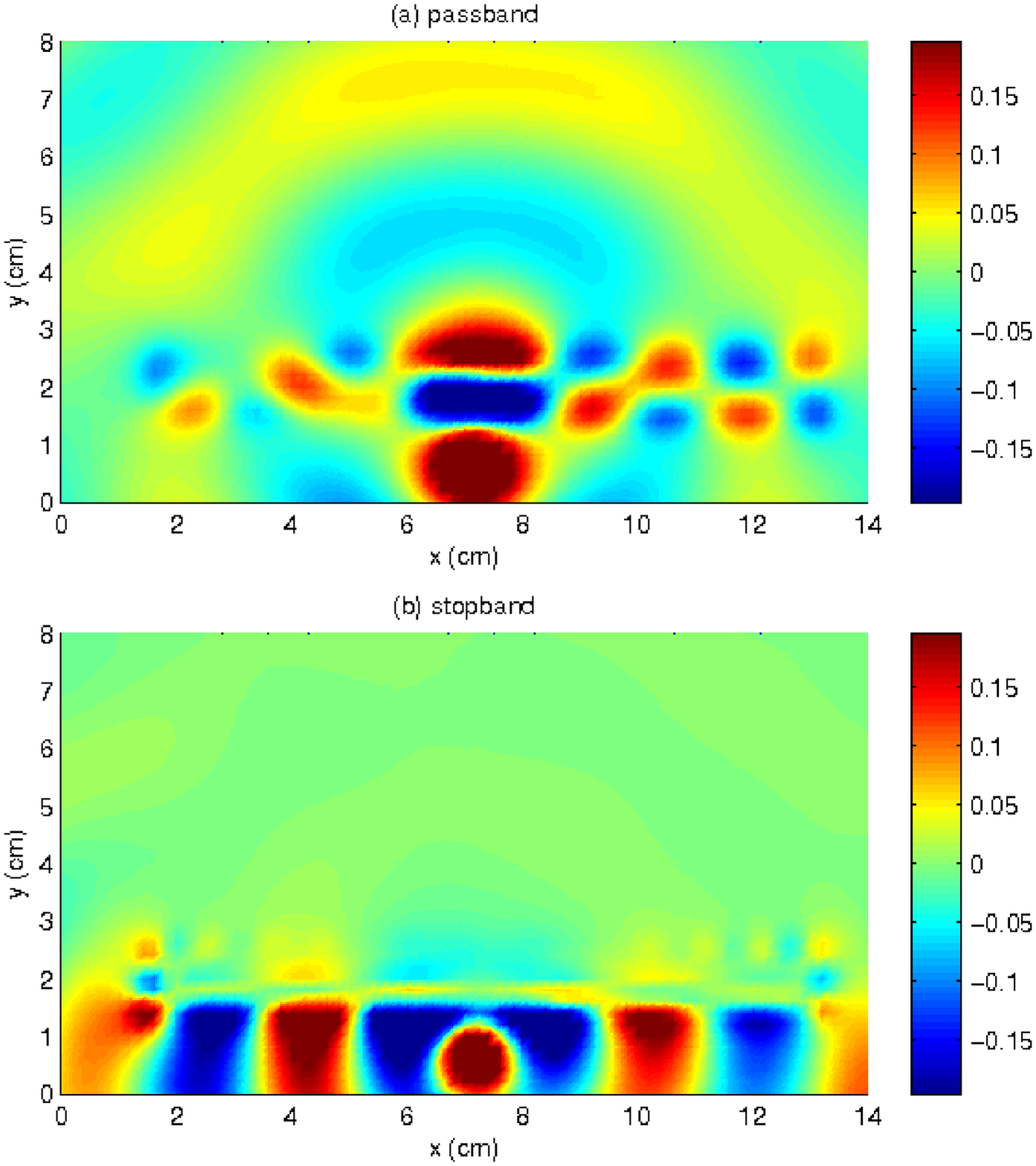}
 \caption{Field scan of the isotropic two-dimensional cut-wire-pairs metamaterial
 (a) in the pass-band at 6.0GHz (Media 1), and (b) in the stop-band at 8.0GHz (Media 2). \label{fig:scan}}
\end{center}
\end{figure}

The two-dimensional response of the structure is examined by exciting it with a point source, and scanning the resultant field distribution using a probe which is flush with the top surface of the waveguide, thus allowing fields inside the metamaterial to be measured.  Figure~\ref{fig:scan} (Media 1 and 2) shows the electric field in the pass-band region (below the stop-band), and in the stop-band.  In the pass-band it can be seen that cylindrical wavefronts propagate beyond the material, albeit with some attenuation, and that there is a complex wave structure within the material due to multiple reflections.  Within the stop-band transmission through the metamaterial is strongly inhibited, and surface waves propagate along the front interface; this is confirmed in the the accompanying animated plots which also incorporate the measured phase information.  This indicates that the negative magnetic response exists for all directions of propagation in the plane of our metamaterial.

For comparison purposes we also show the measured and simulated transmission spectra of the two orthogonal sets of boards used to create the isotropic metamaterial.  In Fig.~\ref{fig:transmission}, the curve labeled ``perpendicular'' corresponds to the boards being perpendicular to the direction of propagation, similar to the one dimensional structures previously reported in the literature.  It can be seen that the isotropic structure has a lower stop-band frequency.

\begin{figure}[htb]
\begin{center}
 \includegraphics[width=0.75\columnwidth]{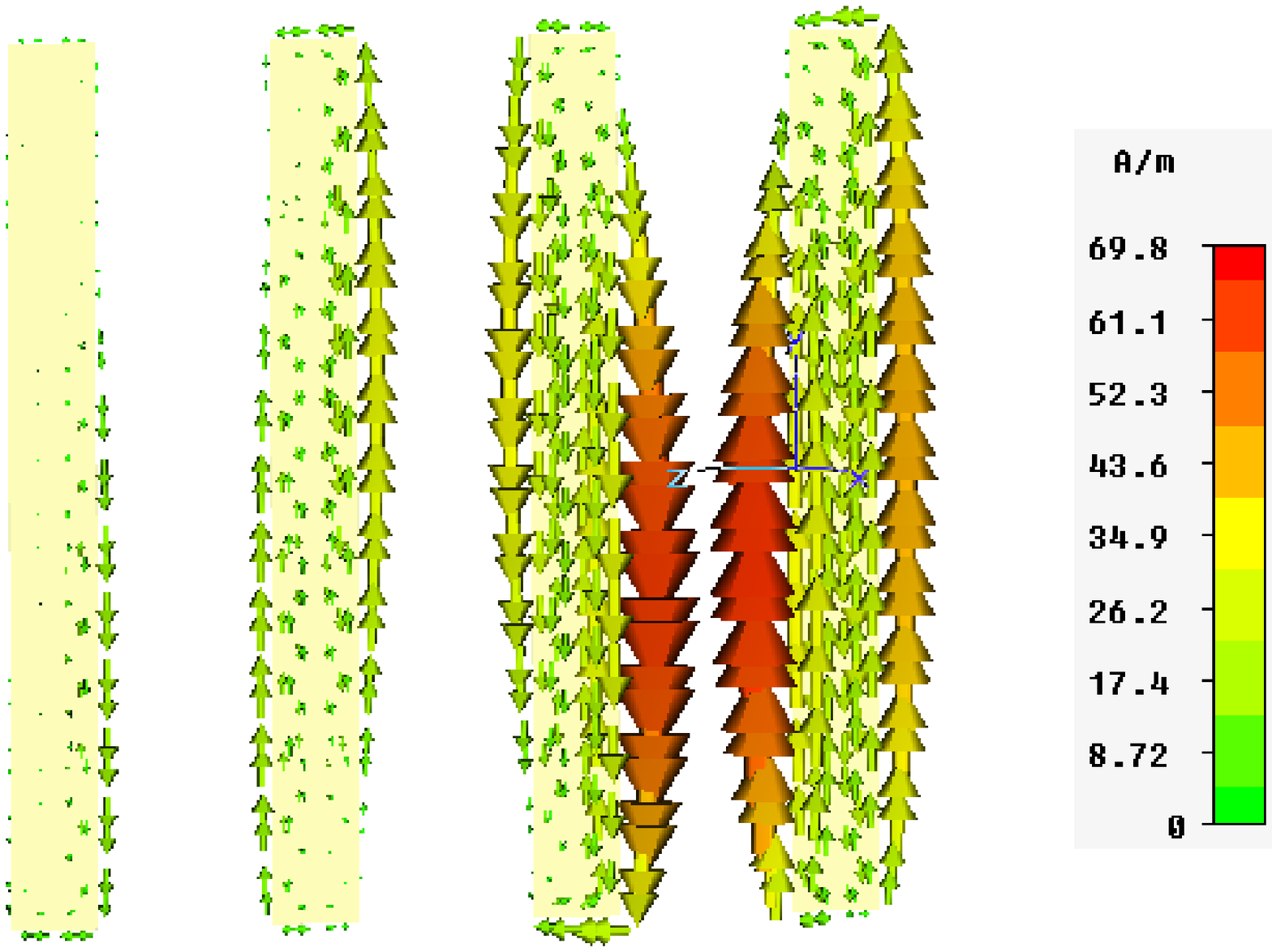}
\caption{Simulated current distribution within the stop-band
at 9GHz, for the boards oriented parallel to the direction of propagation, showing that coupling
between adjacent resonators gives rise to a magnetic response.~\label{fig:currents-par}}
\end{center}
\end{figure}

The curves labeled ``parallel'' correspond to the boards oriented parallel to the direction of propagation.  In contrast to the situation with split-ring resonators, which have a strong magnetic response only along a single axis, \emph{the neighboring pairs of cut-wires in adjacent unit cells can form a current loop to yield a magnetic response}.  The currents corresponding to this case are shown in Fig. \ref{fig:currents-par}, where it can be seen that currents in adjacent pairs are out of phase, and effectively form another cut-wire pair. 

However, this does not result in a clear stop-band, due to the slots in the boards and the FR4 frame required to hold the sample in place.  Simulating this layout without the slots results in a clear magnetic stop-band, as shown by the curve labeled ``parallel (ideal)''. The two resonances of the orthogonal sets of resonators at similar frequencies interact with each other, resulting in an increase in the effective inductance or capacitance of the two-dimensional system.

\begin{figure}[htb]
\begin{center}
 \includegraphics[width=\columnwidth]{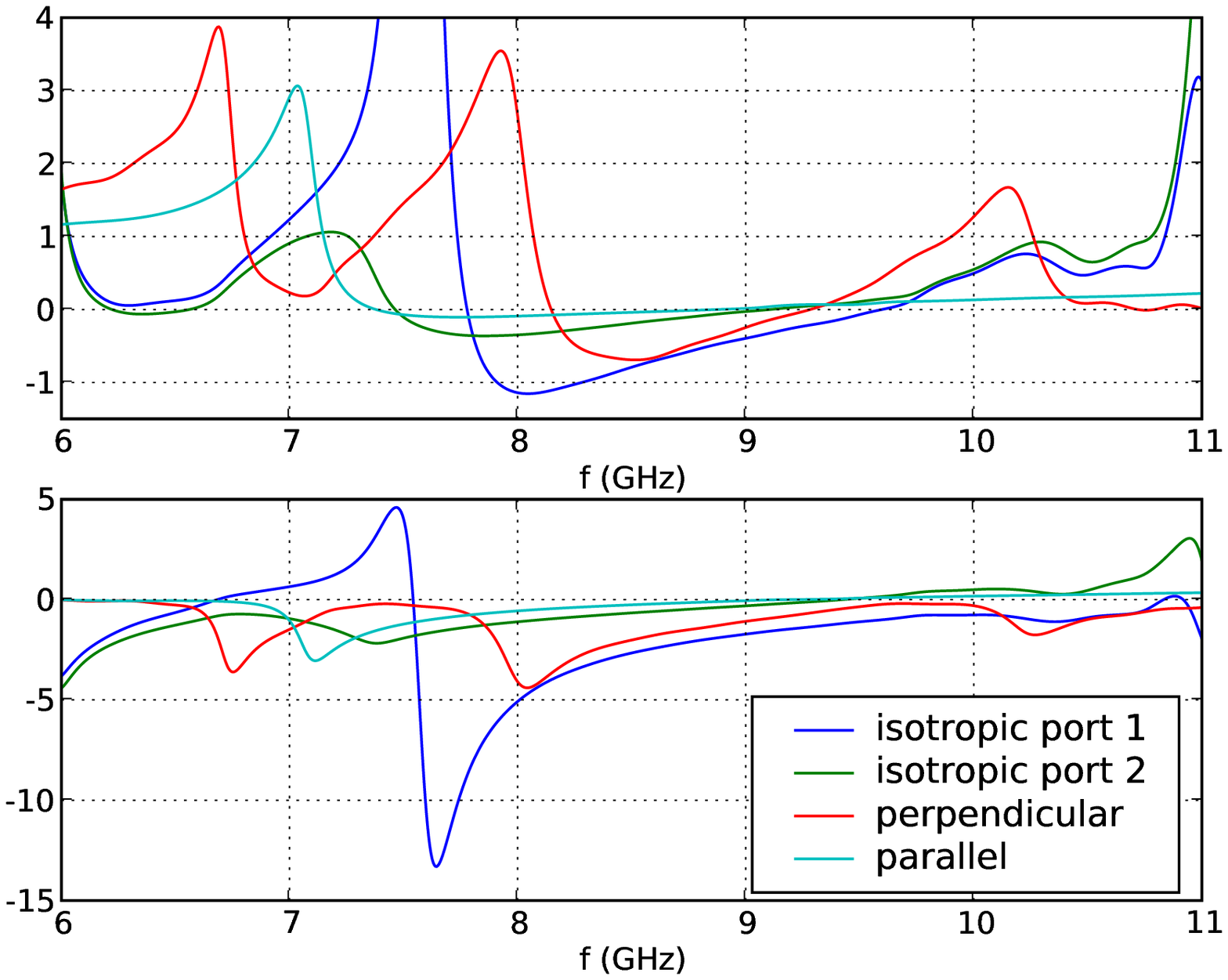}
\caption{The permeability extracted from simulated reflection and transmission data\label{fig:mu}}
\end{center}
\end{figure}

In order to verify the magnetic nature of the response of the metamaterials, we calculate effective parameters from simulated transmission and reflection data, using the method described in Ref.~\cite{Chen2004}. The permeability is plotted in Fig. \ref{fig:mu} for all three structures considered (the ``parallel'' structure does not include the effects of slots), and in all cases a negative magnetic response is observed over some frequency band.  The asymmetry of the unit cell for the isotropic structure results in different effective parameters for excitation from the front and rear~\cite{Smith2005}.  In addition, in the regions of strong resonant response there are regions where the imaginary part of the permeability becomes positive.  This problem is well known, and further discussion can be found e.g. in Ref. \cite{Smith2005,Chen2004}.

For creating a structure scalable beyond microwave frequencies, the use of interlocking boards becomes impractical.  Examining Fig.~\ref{fig:currents}, it is clear that currents within the closest wires (i.e. those not part of the same pair) are in phase.  Thus the four nearest wires could be combined into a single solid conductor, and the boards could be replaced by a continuous dielectric background, with little qualitative change in the response.  We have modeled such a structure and found it shows similar stop-band behavior to the two-dimensional structure discussed above, but with stop-band shifted downwards in frequency by approximately 10\%.  Such a structure could be fabricated in a planar configuration by milling or anisotropic etching of holes which are then filled with metal using a conformal deposition process, followed by lapping.

In conclusion, we have experimentally demonstrated that cut-wire pair metamaterials
can be extended to two dimensions. We have suggested a novel design of cut-wire metamaterials and demonstrated a good agreement between experimentally measured data and the results calculated numerically.  We have observed that, unlike the structures based on split-ring resonators, the nearest-neighbor coupling in cut-wire-pair structures can result in a magnetic stop-band with propagation in the transverse direction. While our experimental results have been obtained for microwaves, we believe our approach will be effective for creating two-dimensional optical metamaterials.  This work was supported by the Australian Research Council.

\end{document}